\documentclass[onecolumn,floatfix,superscriptaddress,showpacs,showkeys,nofootinbib]{revtex4}%
\usepackage{epsfig}
\usepackage{amssymb,latexsym,amsmath}
\newcommand{\eq}[1]{\begin{align} #1 \end{align}}

\usepackage{bm}
\usepackage{amsmath}
\usepackage{graphicx}
\usepackage{subfigure}
\usepackage[colorlinks=true,linktocpage=true,linkcolor=blue,citecolor=blue,allcolors=blue]{hyperref}
\usepackage[usenames,dvipsnames]{color}

\begin{document}

\title{Participant number fluctuations for higher moments of a multiplicity distribution}

\author{Viktor Begun}
 \email{viktor.begun@gmail.com}
 \affiliation{Institute of Physics, Jan Kochanowski University, PL-25406~Kielce, Poland}

%\date{June 20, 2016}

\begin{abstract}
The independent participant model is generalized for skewness and
kurtosis.
The obtained relations allow to calculate the fluctuations of an
arbitrarily high order.
From the comparison with the SPS and the LHC data it is found that
the participants are not nucleons.
The contribution of the participant fluctuations increases with
the order of fluctuations.
The 5\% centrality bins selected for the analysis at the LHC by
ALICE seems to be too large. The fluctuations measures are
dominated by the fluctuations of participants there.
The method to quantify the value of participant number
fluctuations experimentally is proposed.
\end{abstract}

\pacs{}%25.75.-q, 24.60.Ky, 67.85.Hj, 25.75.Ld}

\keywords{wounded nucleons, independent participants, independent
sources, participant fluctuations, scaled variance, skewness,
kurtosis, higher moments}

\maketitle

%%%%%%%%%%%%%%%%%%%%%%%%%%%%%%%%%%%%%%%%%%%%%%%%%%%%%%%%%%%%%%%%%%%%%%%%%%%%%%%%%%%%%%%%%%%%%%%%%%%
%\section{Introduction}
%\label{sect:Intro}

Many observables in high energy collisions scale with the number
of participants - $N_{\rm P}$ - the nucleons that interacted
inelastically during a collision. The total number of charged
particles is proportional to $N_{\rm P}$ in the measured energy
range~\cite{Elias:1978ft,Back:2005hs,Abbas:2013bpa}.
This effect was addressed first in Ref.~\cite{Bialas:1976ed}
within the wounded nucleon model. It is based on the optical limit
of the famous Glauber model~\cite{Glauber-Notes,Glauber:2006gd},
which physically means that the wounded nucleons emit particles
independently from each other. The latter is more general
requirement the same as in the independent source model or
independent participant model. The term participants is the most
commonly used now, therefore it is used later on in this paper.

It was found that the behavior of the scaled variance of a
multiplicity distribution as the function of $N_{\rm P}$ can be
also qualitatively explained by the fluctuations of
participants~\cite{Konchakovski:2005hq,Konchakovski:2007zza}. The
scaled variance is proportional to the second moment of a
multiplicity distribution. The ratio of the fourth to the square
of the second moment is called kurtosis.
The STAR collaboration observes the non-monotonous behavior of the
normalized kurtosis for the net-proton
distribution~\cite{Adamczyk:2013dal,Luo:2015ewa}. This might be an
indication of the critical point of strongly interacting matter,
as higher moments of fluctuations are more sensitive to the
proximity of the QCD critical point~\cite{Stephanov:2008qz}. It is
quite intriguing that the kurtosis has a minimum in the vicinity
of the collision energy where the NA49 and the STAR collaborations
see the famous $K^+/\pi^+$
horn~\cite{Gazdzicki:1998vd,Alt:2007aa,Aggarwal:2010cw,Das:2014qca}.
A significant effort of the NA61/SHINE collaboration, the
successor of the NA49, is going to be devoted to the study of high
order fluctuations.
A most challenging background for these studies seems to be the
fluctuations of nucleon participants, similar to the case with the
scaled variance. These fluctuations are experimentally unavoidable
and, therefore, should be reliably estimated.
One can derive the necessary formulas for the third moment from
the Ref.~\cite{Olszewski:2015xba}, see
also~\cite{Mrowczynski:1999un}. However, it seems that, in spite
of the practical importance, the influence of the fluctuations of
nucleon participants for higher moments was not considered yet.

In the present paper the expressions for the third and the fourth
moments are written explicitly for arbitrary distributions of both
measured particles and the participants. The only assumptions are
that the participants are {\it identical} and {\it independent}. %, the same as in the wounded nucleon model.
The proposed method of calculation allows to derive
straightforwardly the influence of the participant fluctuations
for arbitrarily high moments.
\\

The multiplicity of some particles $N$ created in a collision is
the sum of the contributions from $N_{\rm P}$ participants
 \eq{
 N &~=~ n_1 ~+~ n_2 ~+~\ldots~+~ n_{_{N_{\rm P}}}.
 }
The number of particles $n_i$ from a participant $i$ fluctuates.
If the participants are identical, then the average $\langle n_i
\rangle = \langle n_j \rangle = \langle n_1 \rangle$ and
\eq{\label{N1}
 \langle N\rangle
 ~= \sum_{N_{\rm P}}P(N_{\rm P}) ~\langle~ \sum_{i=1}^{N_{\rm P}}n_i ~\rangle
 ~= \sum_{N_{\rm P}}P(N_{\rm P})~N_{\rm P}~\langle n_1 \rangle
 ~=~ \langle N_{\rm P}\rangle~ \langle n_1\rangle~,
 }
where $P(N_{\rm P})$ is the probability distribution of the
participants number. Similarly
\eq{\label{N2}
 \langle N^2\rangle
 &~= \sum_{N_{\rm P}}P(N_{\rm P})~\langle \left(~\sum_{i=1}^{N_{\rm P}} n_i \right)^2\rangle
  \nonumber\\
 &~= \sum_{N_{\rm P}}P(N_{\rm P})\left(\sum_{i=1}^{N_{\rm P}} \langle n_i^2 \rangle ~+~ \sum_{i\neq j= 1}^{N_{\rm P}} \langle n_in_j \rangle \right)
  \nonumber\\
 &~=~ \langle N_{\rm P}\rangle~ \langle n_1^2\rangle ~+~ \langle N_{\rm P}(N_{\rm P}-1)\rangle~ \langle n_1\rangle^2~,
 }
where the assumption that the participants are independent
$\langle n_i n_j\rangle=\langle n_i\rangle \langle
n_j\rangle=\langle n_1\rangle^2$ is used.
Equations (\ref{N1}) and (\ref{N2}) give the famous formula for
the scaled variance,
 \eq{\label{w}
 \omega ~=~ \frac{\langle N^2\rangle-\langle N\rangle^2}{\langle N\rangle} ~=~ \omega_1 ~+~ \langle n_1\rangle ~ \omega_{\rm P}~,
 }
which is present already in~\cite{Bialas:1976ed}. It is the sum of
the fluctuations from one participant $\omega_1$ and the
fluctuations of participant number $\omega_{\rm P}$ times the mean
multiplicity of particles of interest from one participant
$\langle n_1\rangle$.
Using the multinomial theorem,
 \eq{
 \left(n_1+n_2+\ldots+n_{_{N_{\rm P}}} \right)^k
 ~=~ \sum_{k_1, k_2,\ldots k_{_{N_{\rm P}}}}\frac{k!}{k_1!k_2!\ldots k_{_{N_{\rm P}}}!}~n_1^{k_1}n_2^{k_2}\ldots n_{_{N_{\rm P}}}^{k_{_{N_{\rm P}}}}~
     \delta\left(k-\sum_{i=1}^{N_{\rm P}} k_i\right)~,
 }
where $\delta$ is the Kronecker delta function, one can obtain
arbitrarily high moment in the model of independent participants.
For the third and the fourth moments one has:
 \eq{
 \langle N^3\rangle
 &~=~~\,\,\langle N_{\rm P}\rangle ~ \langle n_1^3\rangle
  \nonumber \\
 &~+~  3\,\langle N_{\rm P}(N_{\rm P}-1)\rangle ~ \langle n_1^2\rangle \langle n_1\rangle
 \nonumber \\
 &~+~~\,\,\langle N_{\rm P}(N_{\rm P}-1)(N_{\rm P}-2)\rangle ~ \langle n_1\rangle^3~,
 \\
  \langle N^4\rangle
 &~=~~\,\langle N_{\rm P}\rangle ~ \langle n_1^4\rangle
  \nonumber \\
 &~+~  4\,\langle N_{\rm P}(N_{\rm P}-1)\rangle ~ \langle n_1^3\rangle\langle n_1\rangle
  \nonumber \\
 &~+~  3\,\langle N_{\rm P}(N_{\rm P}-1)\rangle ~ \langle n_1^2\rangle^2
  \nonumber \\
 &~+~  6\,\langle N_{\rm P}(N_{\rm P}-1)(N_{\rm P}-2)\rangle ~ \langle n_1^2 \rangle ~ \langle n_1\rangle^2
 \nonumber \\
 &~+~~\,\,\langle N_{\rm P}(N_{\rm P}-1)(N_{\rm P}-2)(N_{\rm P}-3)\rangle ~ \langle n_1\rangle^4~.
 \label{N4}
 }
The coefficients in front of the $\langle n_1^{k_i}\rangle^{k_j}$
terms are given by the product of the multinomial coefficient, the
number of permutations $\frac{N_{\rm P}!}{(N_{\rm P}-k_i)!}$, and
the additional degeneracy factor that appears due to the fact that
the emitted particles are indistinguishable. For example, the
factor before $\langle n_1\rangle^4$ in (\ref{N4}) is equal to the
multinomial coefficient $\frac{4!}{1!1!1!1!}=4!$ times the number
of ways to pick up four different participants $\frac{N_{\rm
P}!}{(N_{\rm P}-4)!}$, divided by the degeneracy factor $4!$ due
to the replacement $\langle n_i\rangle\langle n_j\rangle\langle
n_k\rangle\langle n_l\rangle = \langle n_1\rangle^4$. The
coefficient in front of $\langle n_1^2\rangle\langle n_1\rangle^2$
in (\ref{N4}) is equal to $\frac{4!}{2!1!1!0!}=12$ times
$\frac{N_{\rm P}!}{(N_{\rm P}-3)!}$, divided by $2!$ due to
$\langle n_i\rangle\langle n_j\rangle=\langle n_1\rangle^2$, etc..
The sum of all the coefficients for $\langle
n_1^{k_i}\rangle^{k_j}=1$ before the averaging over participants
gives $N_{\rm P}^k$, which can be used for a quick check. The
formulas for higher moments can be derived in the similar way.
The raw moments $\langle N^k\rangle$ are directly related to
central moments of a distribution $P(N)$
 \eq{\label{mk}
 m_k~=~\sum(N-\langle N\rangle)^kP(N)~.
 }
The second, the third, and the fourth moments in the model of
independent participants equal to:
 \eq{
 m_2
 &~=~ \langle N_{\rm P}\rangle\, m_2^1 ~+~ \langle n_1\rangle^2\, m_2^{\rm P}~,
%  ~=~ \langle N\rangle\left(\omega_1+\langle n_1\rangle\omega_{\rm P}\right)
 \\
 m_3
 &~=~ \langle N_{\rm P}\rangle\, m_3^1 ~+~ \langle n_1\rangle^3\, m_3^{\rm P} ~+~ 3\,\langle n_1\rangle\, m_2^{\rm P}\,m_2^1~,
 \\
% m_4
% &~=~ \langle N_{\rm P}\rangle\,\left[\,
%      m_4^n
%  ~+ \left[\, \langle N_{\rm P}\rangle -1\,\right]
%      3(m_2^n)^2
%  ~+~ 6 \langle n_1\rangle^2
%      m_2^n m_2^{\rm P}
%      \,\right]
%  \nonumber \\
% &~~~~+~ m_2^{\rm P}
%         \left[\,
%      4 \langle n_1\rangle\,
%         m_3^n
%  ~+~ 3 (m_2^n)^2
%        \,\right]
%  ~+~ 6 \langle n_1\rangle^2\,
%         m_2^n\, m_3^{\rm P}
%  ~+~   \langle n_1\rangle^4\,
%         m_4^{\rm P}
% \nonumber \\
  m_4
 &~=~ \langle N_{\rm P}\rangle\, (m_4^1 - 3(m_2^1)^2)
  ~+~ 3\,m_2^{\rm P}(m_2^1)^2
  ~+~ 4\,\langle n_1\rangle\,   m_2^{\rm P}\, m_3^1
  ~+~ 6\,\langle n_1\rangle^2\, m_3^{\rm P}\, m_2^1
 \nonumber \\
 &~+~  \langle n_1\rangle^4\, \left[\, m_4^{\rm P} - 3(m_2^{\rm P})^2\,\right]
 \,+~ 3\,(m_2)^2~,
 }
where $m_k^1$ and $m_k^{\rm P}$ are defined the same as $m_k$
(\ref{mk}) for the distribution of particles produced by one
source $P(n_1)$ and for the distribution of participants $P(N_{\rm
P})$.

The combination of central moments gives the scaled variance, the
normalized skewness, and the normalized kurtosis:
 \eq{
 \omega           ~=~ \frac{m_2}{\langle N\rangle}~,&&
 S\,\sigma        ~=~ \frac{m_3}{m_2}~,&&
 \kappa\,\sigma^2 ~=~ \frac{m_4}{m_2}~-~3\,m_2~,&&
 \text{where}~~\sigma^2=m_2~.
 }
They describe the width, the asymmetry, and the sharpness of a
distribution with a single maximum, correspondingly. Skewness and
kurtosis are much more sensitive to the properties of a
multiplicity distribution. A Poisson distribution has
$\omega=S\sigma=\kappa\sigma^2=1$ for the same mean multiplicity,
while $\omega$ is a free parameter and $S\sigma=\kappa\sigma^2=0$
for Normal (Gauss) distribution.
In the independent participant model the normalized skewness
equals
 \eq{\label{Ssig}
 S\,\sigma ~=~ \frac{\omega_1~S_1\sigma_1
 ~+~ \langle n_1\rangle~\omega_{\rm P}\left[~3\,\omega_1 ~+~ \langle n_1\rangle\,S_{\rm P}\,\sigma_{\rm P}~\right] }
        {\omega_1 ~+~ \langle n_1\rangle ~ \omega_{\rm P}}~,
 }
and normalized kurtosis:
 \eq{\label{ks2}
 \kappa\,\sigma^2
 ~=~ \frac{\omega_1~\kappa_1\sigma_1^2
        ~+~\omega_{\rm P}~
            \left[~\langle n_1\rangle^3~\kappa_{\rm P}\,\sigma_{\rm P}^2
        ~+~\langle n_1\rangle~\omega_1
           \left(~
           3\,\omega_1
        ~+~4\,S_1\sigma_1
        ~+~6\,\langle n_1\rangle\,S_{\rm P}\,\sigma_{\rm P}
           ~\right)~\right]}
     {\omega_1 ~+~ \langle n_1\rangle ~ \omega_{\rm P}}~.
 }

Scaled variance, skewness and kurtosis depend crucially on the
strength of participant fluctuation $\omega_{\rm P}$. If it is
zero, then the information about participants is left in the mean
multiplicity, but is cancelled in fluctuations:
 \eq{
 \langle N\rangle ~=~ \langle N_{\rm P}\rangle\,\langle n_1\rangle~,&&
  \omega          ~=~ \omega_1~,&&
  S\,\sigma       ~=~ S_1\,\sigma_1~,&&
 \kappa\,\sigma^2 ~=~ \kappa_1\,\sigma_1^2~,&&\text{for}~~\omega_{\rm P}~=~0~,
 }
so that one observes the fluctuations from one source. It is a
desired situation, because participant fluctuations are mainly
driven by the uncertainty of the centrality determination. They
may mimic or hide the QCD critical point and any other signal. The
fluctuations of participants seem to be unavoidable, because one
always has a finite centrality window in experiment. If this
window is too narrow, then one may cut also the fluctuations from
one source. Therefore, one should find the balance between
fluctuations of participants $\omega_{\rm P}$, the number and
fluctuations of particles from one participant $\omega_1/\langle
n_1\rangle$.

For small fluctuations of the participants, $\omega_{\rm
P},\,S_{\rm P}\sigma_{\rm P}\ll\omega_1/\langle n_1\rangle$ and
$\kappa_{\rm P}\sigma_{\rm P}^2\ll\omega_1^2/\langle
n_1\rangle^2$, one obtains:
 \eq{\label{SsigKs2-1}
 \omega    ~\simeq~ \omega_1~,
 &&
 S\,\sigma ~\simeq~ S_1\,\sigma_1 ~+~ 3\,\langle n_1\rangle~\omega_{\rm P} ~,
 &&
 \kappa\,\sigma^2 ~\simeq~ \kappa_1\,\sigma_1^2 ~+~ \langle n_1\rangle~\omega_{\rm P}\,(3\,\omega_1+4\,S_1\,\sigma_1)~,
 }
i.e., the scaled variance is determined by the fluctuations from
one participant, however skewness and kurtosis further depend on
how large is the product $\langle n_1\rangle~\omega_{\rm P}$
compared to the skewness and kurtosis for one source.
The fluctuations from one source should be large close to critical
point or phase transition. For example, all moments higher than
$k>2$ diverge at Bose-Einstein condensation~\cite{Begun:2016cva},
which is the third order phase transition.

For large enough fluctuations of the participants, $\omega_{\rm
P}\gg\omega_1/\langle n_1\rangle$, $\omega_{\rm
P}\gg\kappa_1\sigma_1^2/(\langle n_1\rangle\,\omega_1)$, and
$S_{\rm P}\sigma_{\rm P}\gg\omega_1/\langle n_1\rangle$, $S_{\rm
P}\sigma_{\rm P}\gg S_1\sigma_1/\langle n_1\rangle$ one finds:
 \eq{\label{SsigKs2-P}
 \omega    ~\simeq~ \langle n_1\rangle\, \omega_{\rm P}~,
 &&
 S\,\sigma ~\simeq~ \langle n_1\rangle ~ S_{\rm P}\,\sigma_{\rm P} ~+~ 3\,\omega_1~,
 &&
 \kappa\,\sigma^2 ~\simeq~ \langle n_1\rangle^2\, \kappa_{\rm P}\,\sigma_{\rm P}^2 ~+~
 6\,\langle n_1\rangle\,\omega_1\,S_{\rm P}\,\sigma_{\rm P}~,
 }
i.e., the observed fluctuations are determined mainly by the
fluctuations of the participants. Note the $\langle n_1\rangle$
and $\langle n_1\rangle^2$ multipliers in front of scaled
variance, skewness and kurtosis from participants in
(\ref{SsigKs2-P}). For large energies $\langle n_1\rangle$ grows
fast and leads to the domination of participant fluctuations for
high moments even for relatively small $\omega_{\rm P}$, $S_{\rm
P}\,\sigma_{\rm P}$ and $\kappa_{\rm P}\,\sigma_{\rm P}^2$.
The participant fluctuations are rather large in a standard
centrality interval. A finer centrality
selection~\cite{Begun:2006uu} or(and) special variables should be
used to cancel the fluctuations of
participants~\cite{Gorenstein:2011vq,Begun:2012wq,Begun:2014boa}.

The experimental information on participant fluctuations is quite
ambiguous. The behavior of the scaled variance of a multiplicity
distribution in nucleus-nucleus (A+A) collisions as the function
of $N_{\rm P}$ was qualitatively explained by the fluctuations of
participants both at SPS and at
RHIC~\cite{Konchakovski:2005hq,Konchakovski:2007zza}. However,
more recent data of NA49 and
NA61/SHINE~\cite{Lungwitz:2006cx,Rybczynski-ISMD2013,Aduszkiewicz:2015jna}
show that
 \eq{\label{wNA49}
 \omega_{\rm Pb+Pb}^{\rm ch}~<~\omega_{\rm p+Pb}^{\rm ch}~<~\omega_{\rm p+p}^{\rm ch}\qquad \text{at SPS}~,
 }
while one would expect the opposite dependence from the
participant model. Using Eq.~(\ref{w}) one obtains
 \eq{\label{wPart}
 \omega_{\rm Pb+Pb}^{\rm ch}~=~\omega_1~+~\langle n_{\rm Pb+Pb}^{\rm ch}\rangle~\omega_{\rm
 P}~,\qquad
 \text{where}~~\langle n_{\rm Pb+Pb}^{\rm ch}\rangle~=~\frac{\langle N_{\rm Pb+Pb}^{\rm ch}\rangle}{\langle N_{\rm P}\rangle}~,
 }
and $\langle n_{\rm Pb+Pb}^{\rm ch}\rangle$ is the number of
charged particles per participant. One can see from
Eqs.~(\ref{wNA49}) and (\ref{wPart}) that the fluctuations in
Pb+Pb and p+Pb can not be constructed from fluctuations of p+p.
Both $\langle n_{\rm Pb+Pb}^{\rm ch}\rangle$ and $\omega_{\rm P}$
are positive, therefore, if $\omega_1=\omega_{\rm p+p}^{\rm ch}$
in (\ref{wPart}), then fluctuations of participants, $\omega_{\rm
P}$, must be negative in this case. It is impossible, since
$\omega$ is positive by definition. However, the fluctuations in
p+p and in Pb+Pb are similar at SPS, therefore the
relation~(\ref{wNA49}) may be attributed to a combination of some
other effects.

The situation should be clear at the LHC, because p+p roughly
follow the KNO scaling, which leads to $\omega_{\rm p+p}^{\rm
ch}\sim\langle N^{\rm ch}_{\rm p+p}\rangle$ and a fast rise of
fluctuations with increasing the energy of the collision,
$\sqrt{s_{NN}}$, while for A+A a weaker dependence of fluctuations
with energy is expected~\cite{Heiselberg:2000fk}. The ALICE
collaboration has published the results for fluctuations of
charged particles $\omega^{\rm ch}$ in Pb+Pb collisions within the
$|\eta|<0.8$ rapidity range.
Their comparison with the AMPT and HIJING string transport models
shows different fluctuations and the different dependence on
$\langle N_{\rm P}\rangle$ than in the
experiment~\cite{Mukherjee:2016hrj}.

%Transport models are the participants model by construction,
%therefore, this discrepancy may be the indication of the
%inapplicability of participant model for fluctuations at the LHC,
%which supports the finding~(\ref{wNA49}) at the SPS.

Instead of running a transport code one may solve the inverse
task. Namely, determine how large should be the fluctuations of
the participants in order to describe the data, assuming different
fluctuations of the participants.
The CMS and the ALICE collaborations have published the data for
fluctuations in p+p~\cite{Khachatryan:2010nk,Adam:2015gka}, as
well as the rapidity distributions of charged particles, and the
number of participants at different centralities in Pb+Pb at
$\sqrt{s_{NN}}=2.76$~TeV~\cite{Abbas:2013bpa,Adam:2015kda}.
Therefore, one can check whether the fluctuations in Pb+Pb is the
sum of the fluctuations in p+p and the fluctuations of
participants.
One should take the measured fluctuations in Pb+Pb, $\omega_{\rm
Pb+Pb}^{\rm ch~Acc}$, from ALICE~\cite{Mukherjee:2016hrj}.
Calculate the rate of how many charged particles are accepted
within their rapidity window, $|\eta|<0.8$, with respect to the
number of charged particles in the full rapidity $q=\langle N_{\rm
Pb+Pb}^{\rm ch}\rangle\big|_{|\eta|<0.8}\,/\langle N_{\rm
Pb+Pb}^{\rm ch}\rangle$. Then one should use the well known
acceptance formula for scaled variance, see e.g.
Ref.~\cite{Begun:2004gs},
 \eq{\label{wAcc-q}
 \omega^{\rm Acc}~=~1\,-\,q\,+\,q\,\omega~,
 }
to reconstruct the fluctuations in the full rapidity range,
$\omega=\omega_{\rm Pb+Pb}^{\rm ch}$, then use Eq.~(\ref{wPart}),
and find
 \eq{\label{wAcc}
 \omega_{\rm Pb+Pb}^{\rm ch~Acc}
 ~=~ 1~-~q ~+~ q~\omega_{\rm Pb+Pb}^{\rm ch}
 ~=~ \omega_1^{\rm Acc} ~+~ \langle n_{\rm Pb+Pb}^{\rm ch~Acc}\rangle~\omega_{\rm P},
 }
where
%
% \eq{
$\omega_1^{\rm Acc}=1-q+q\,\omega_1$ and $\langle n_{\rm
Pb+Pb}^{\rm ch~Acc}\rangle=q\,\langle n_{\rm Pb+Pb}^{\rm
ch}\rangle$.
% }
%
The fluctuations in p+p equal to $\omega_{\rm p+p}^{\rm
Acc}\simeq4.6,~8.46,~11.36,~13.74$ in the rapidity intervals
$|\Delta\eta|=0.5,~1.,~1.5,~2.4$, correspondingly, therefore,
 \eq{
 \omega_1^{\rm Acc}=~\omega_{\rm p+p}^{\rm Acc}\big|_{|\Delta\eta|<0.8}~\simeq~ 7. ~>~ \omega_{\rm Pb+Pb}^{\rm ch~Acc}~\simeq~3.~,
 }
and the fluctuations of the participants are negative in
(\ref{wAcc}), similar to that at the SPS~(\ref{wNA49}).
The acceptance $q\simeq 15\%$ in Pb+Pb at ALICE. It gives
$\omega_{\rm p+p}=(\omega_{\rm p+p}^{\rm
Acc}\big|_{|\Delta\eta|<0.8}-1+q)/q\simeq41$ for the whole
acceptance. One may argue that some processes may damp the
fluctuations from one participant in Pb+Pb compared to p+p.
Let us pick up some numbers in order to quantify a possible
outcome and consider three cases.

First, the fluctuations from one source equal to the maximal
measured fluctuations in p+p, $\omega_1=\omega_{\rm p+p}^{\rm
ch}\big|_{|\eta|<2.4}=13.74$. Let's call this case 'Maximal'.

Second, the fluctuations from one source are Poisson-like,
$\omega_1=\omega_{\rm Poisson}=1$, called 'Poisson'.
For these two cases we know all the terms in Eq.~(\ref{wAcc}),
except for $\omega_{\rm P}$, which is calculated from
(\ref{wAcc}).

Third case -- we do not know the fluctuations from one source, but
we know that the fluctuations of participants are of the order of
unity, $\omega_{\rm P}=1$, as in HIJING and AMPT in
Ref.~\cite{Mukherjee:2016hrj}, and then calculate $\omega_1$ from
(\ref{wAcc}). This case is called 'Transport'.

\begin{figure}
%\centering
\includegraphics[width=0.49\textwidth]{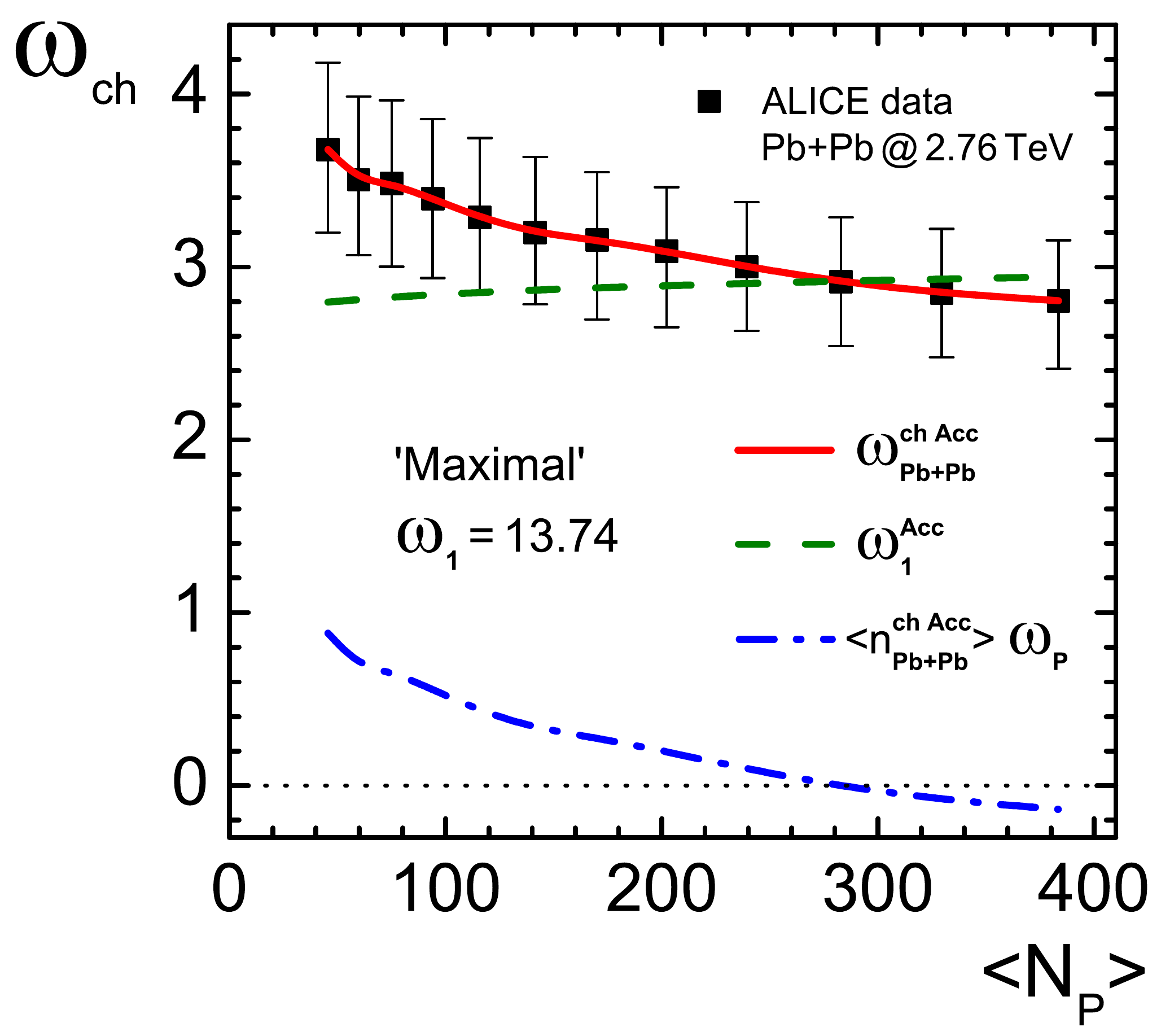}~~
\includegraphics[width=0.49\textwidth]{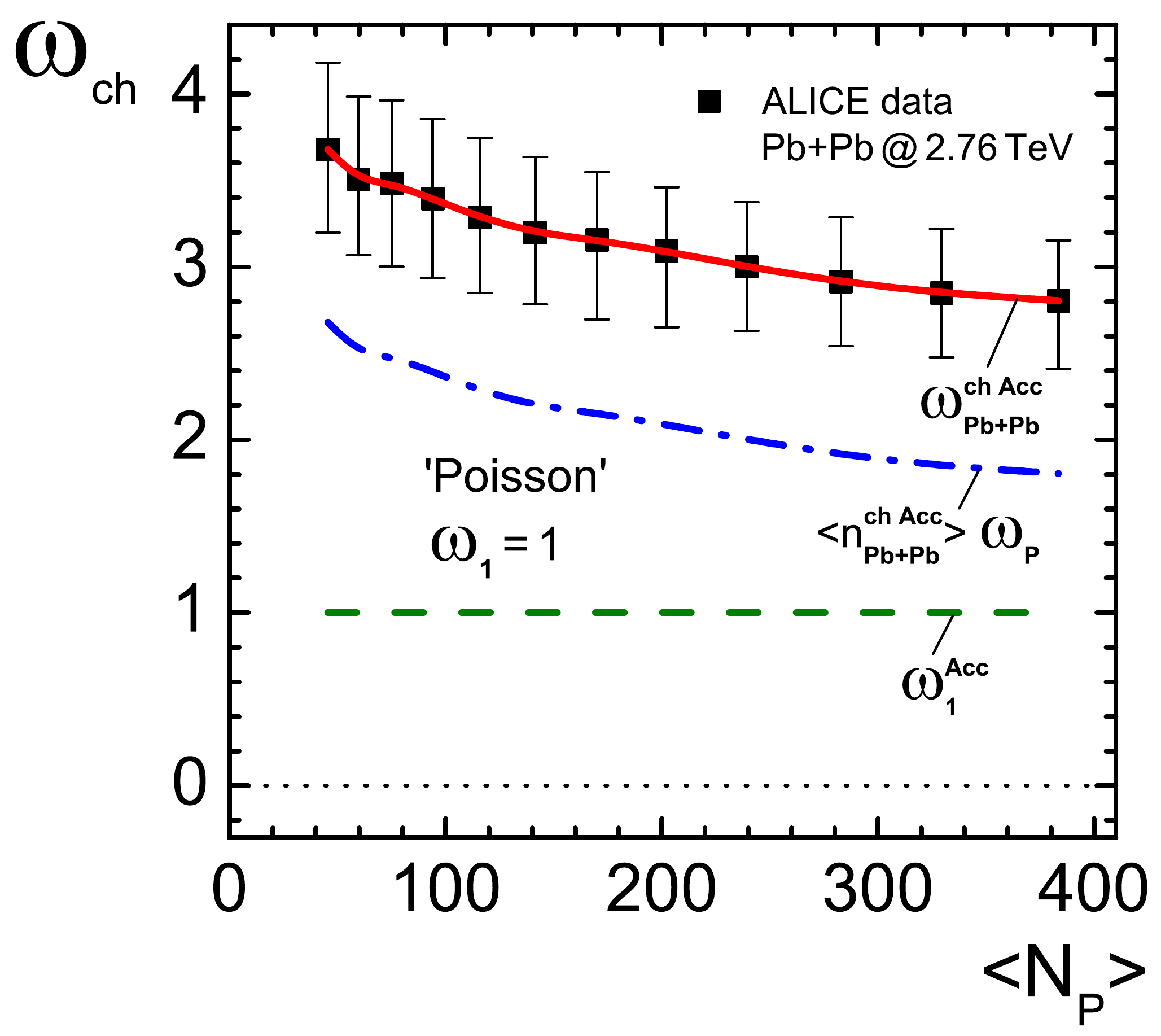}
\caption{The decomposition of experimentally measured fluctuations
of charged particles $\omega_{\rm Pb+Pb}^{\rm ch~Acc}$ as the
function of the number of participants $\langle N_{\rm
P}\rangle$~\cite{Mukherjee:2016hrj} on the fluctuations due to the
fluctuations of participants $\langle n_{\rm Pb+Pb}^{\rm
ch~Acc}\rangle~\omega_{\rm P}$ and due to the fluctuations from
one participant $\omega_1^{\rm Acc}$ in the measured acceptance,
assuming some value of the fluctuations from one participant
$\omega_1$ in the full acceptance.}\label{fig-1}
\end{figure}

\begin{figure}
%\centering
\includegraphics[width=0.49\textwidth]{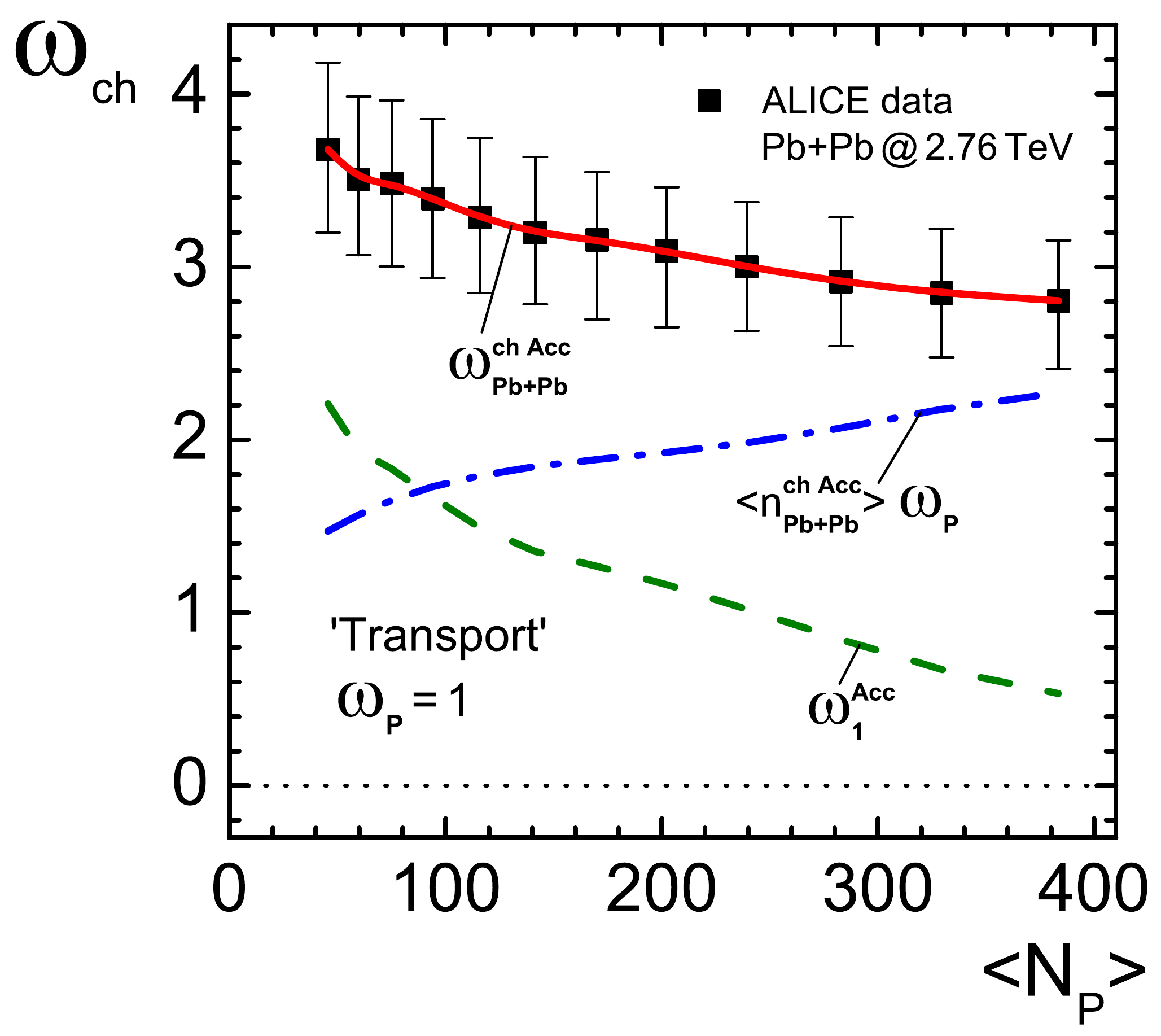}~~
\includegraphics[width=0.49\textwidth]{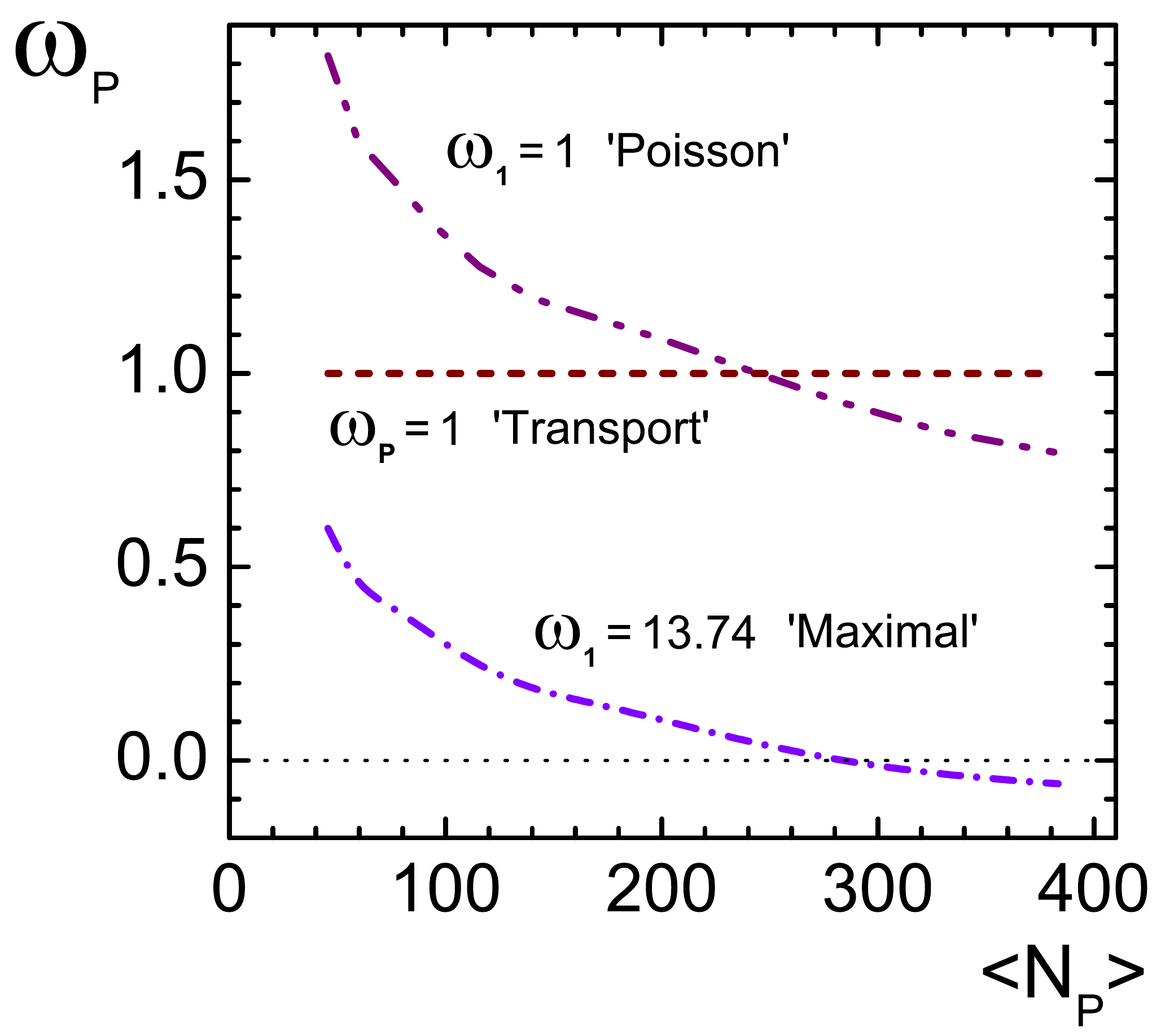}
\caption{Left: The same as Fig.~\ref{fig-1} for the fluctuations
of participants $\omega_{\rm P}=1$. Right: The extracted
fluctuations of participants, assuming some value of the
fluctuations from one source $\omega_1$ in the full
acceptance.}\label{fig-2}
\end{figure}

The results are shown in Figs.~\ref{fig-1}~and~\ref{fig-2}.
The continuous lines in Fig.~\ref{fig-1} and in
Fig.~\ref{fig-2}~left are the fit of the ALICE data for
$\omega_{\rm Pb+Pb}^{\rm ch~Acc}$. They go through the data points
by definition. The dashed and dash-dotted lines is the
decomposition of $\omega_{\rm Pb+Pb}^{\rm ch~Acc}$ into two parts
according to Eq.~(\ref{wAcc}), right.
The corresponding fluctuations of the participants are shown in
Fig.~\ref{fig-2}~right.

The acceptance slowly grows with centrality in the $2.76$~TeV
Pb+Pb collisions at the LHC. Therefore, the 'Maximal' fluctuations
from one source, $\omega_1^{\rm Acc}$, also grow, while the
fluctuations of participants must decrease fast, because the total
fluctuations decrease, see Eq.~(\ref{wAcc}).
The 'Maximal' fluctuations from one source are above the
experimental measurements for large $\langle N_{\rm P}\rangle$,
therefore, the fluctuations of participants, $\omega_{\rm P}$,
become negative, which is forbidden by the definition of $\omega$.
The absolute value of participant fluctuations is small, so that
the measures are done in the 'good' limit (\ref{SsigKs2-1}).
However, $\omega_{\rm P}$ is too small for the LHC, and even
smaller than at RHIC, compare the dash-dotted line in
Fig.~\ref{fig-2} with Fig.~1 from~\cite{Konchakovski:2007zza}.

For 'Poisson' fluctuations from one source the acceptance
dependence is cancelled, $\omega_1^{\rm Acc}=\omega_1=1$ according
to Eq.~(\ref{wAcc-q}), and the fluctuations of participants are
similar to that at RHIC.
However, one expects a strong growth of the participant
fluctuations with energy from transport
models~\cite{Begun:2012wq}.
Moreover, the measurements at ALICE are done in the 'bad' limit,
when all the measures are determined by the fluctuations of the
participants, see Eq.~(\ref{SsigKs2-P}).

The 'Transport' case is in between the 'Maximal' and the
'Poisson', closer to the 'Poisson'. The measurement are done in
the 'bad' limit (\ref{SsigKs2-P}), when the fluctuations of
participants determine the results.
\\

One may conclude that the independent participant model can
describe fluctuations of charged particles in Pb+Pb at the LHC
only if the fluctuations from one participant, $\omega_1$, are
much smaller than the fluctuations of charged particles in p+p
reactions.
Therefore, the participants are not nucleons.

If the fluctuations of participants are larger then Poisson,
$\omega_{\rm P}\geq1$, moreover, if they are as large as predicted
by transport models, then the 5\% centrality bins selected for the
analysis at the LHC by ALICE are too large. In this case the
fluctuations measures are dominated by the fluctuations of
participants and by the corresponding experimental limitations,
like the uncertainty in the centrality determination.

There are many ways to look for a possible solution. The
participants can be quarks, then the number of particles from one
source, $\langle n_{\rm Pb+Pb}^{\rm ch~Acc}\rangle$, reduces three
times, since there are three quarks in each nucleon. It leads to
the increase of the $\omega_{\rm P}$ three times, in order to keep
the same value of the product $\langle n_{\rm Pb+Pb}^{\rm
ch~Acc}\rangle\,\omega_{\rm P}$ in~(\ref{wAcc}).
Another possibility is that the sources are not identical and/or
strongly correlated. The examination of these possibilities
requires further theoretical studies and more data.

%Strictly speaking, we do not know how large are either the
%fluctuations from one source (participant) or the fluctuations of
%participants.
%

One should check experimentally whether participant model works
for fluctuation, eliminate the fluctuations of participants, and
obtain the fluctuations from one source.
In order to do that, one should consider the most central
collisions, reduce the centrality window, and check how the
fluctuations change, taking, let say, $c=0-20\%$, $c=0-15\%$,
$c=0-10\%$, $c=0-5\%$, $c=0-2.5\%$, $c=0-1\%$, etc..
If the participant model works, then one would expect a fast
decrease of the fluctuations due to the decrease of the
participant fluctuations. The decrease should slow down at some
centrality, which is narrow enough, so that the participant
fluctuations do not contribute. If the remaining fluctuations are
not already Poisson-like due to very small acceptance, then these
are the fluctuations from one source.

It seems that the amount of participant fluctuations should be
determined before measurements of the higher moments, because
participants fluctuations may be strong enough to mimic or hide
any other effect.

%%%%%%%%%%%%%%%%%%%%%%%%%%%%%%%%%%%%%%%%%%%%%%%%%%%%%%%%%%%%%%%%%%%%%%%%%%%%%%%%%%%%%%%%%
%
%\section{Conclusions}\label{sect:Concl}
%
%%%%%%%%%%%%%%%%%%%%%%%%%%%%%%%%%%%%%%%%%%%%%%%%%%%%%%%%%%%%%%%%%%%%%%%%%%%%%%%%%%%%%%%%%

%%%%%%%%%%%%%%%%%%%%%%%%%%%%%%%%%%%%%%%%%%%%%%%%%%%%%%%%%%%%%%%%%%%%%%%%%%%%%%%%%%%%%%%%%%%%%%%%%%%
\acknowledgments
%%%%%%%%%%%%%%%%%%%%%%%%%%%%%%%%%%%%%%%%%%%%%%%%%%%%%%%%%%%%%%%%%%%%%%%%%%%%%%%%%%%%%%%%%%%%%%%%%%%

The author thanks to N.~Xu for encouraging discussions at
WPCF~2015 and CPOD~2016 conferences, and to M.~I.~Gorenstein,
W.~Broniowski, V.~Koch, M.~Praszalowicz, A.~Rustamov,
M.~Rybczynski, and I.~Selyuzhenkov, for fruitful comments and
suggestions.
This work was supported by Polish National Science Center grant
No. DEC-2012/06/A/ST2/00390.

%%%%%%%%%%%%%%%%%%%%%%%%%%%%%%%%%%%%%%%%%%%%%%%%%%%%%%%%%%%%%%%%%%%%%%%%%%%%%%%%%%%%%%%%%%%%%%%%%%%
%\bibliographystyle{plain}
\bibliographystyle{h-physrev}
\bibliography{Npart}%{}
%\bibliographystyle{plain}
%%%%%%%%%%%%%%%%%%%%%%%%%%%%%%%%%%%%%%%%%%%%%%%%%%%%%%%%%%%%%%%%%%%%%%%%%%%%%%%%%%%%%%%%%%%%%%%%%%%

\end{document}